\documentclass[twocolumn,prb,superscriptaddress]{revtex4}
\usepackage{graphicx,float}
\usepackage{latexsym}
\usepackage{epstopdf}
\usepackage{amssymb,amsmath}
\usepackage{color}
\usepackage{multirow}
\usepackage{hyperref}
\usepackage{graphicx}
\makeatletter
\let\Hy@linktoc\Hy@linktoc@none
\makeatother
\usepackage{verbatim}
\usepackage{bm}
\begin{document}
\title{Scale-invariant puddles in Graphene: Geometric properties of electron-hole distribution at the Dirac point}
\author{M. N. Najafi*}
\affiliation{Department of Physics, University of Mohaghegh Ardabili, P.O. Box 179, Ardabil, Iran}
\author{M. Ghasemi Nezhadhaghighi}
\affiliation{Department of Physics, College of Science, Shiraz University, Shiraz 71454, Iran}
\begin{abstract}
We characterize the carrier density profile of the ground state of graphene in the presence of particle-particle interaction and random charged impurity for zero gate voltage. We provide detailed analysis on the resulting spatially inhomogeneous electron gas taking into account the particle-particle interaction and the remote coulomb disorder on an equal footing within the Thomas-Fermi-Dirac theory. We present some general features of the carrier density probability measure of the graphene sheet. We also show that, when viewed as a random surface, the resulting electron-hole puddles at zero chemical potential show peculiar self-similar statistical properties. Although the disorder potential is chosen to be Gaussian, we show that the charge field is non-Gaussian with unusual Kondev relations which can be regarded as a new class of two-dimensional (2D) random-field surfaces. 
\end{abstract}
\maketitle
\section{Introduction}
Graphene as a newly realized 2D electron system can be described at low energies by massless Dirac-Fermion model, whose unusual properties has made it as a subject of intence theoretical and experimental research. Many of these studies are still based on idealized models which neglect the effect of disorder and particle-particle interactions. The understanding of the origin and effects of extrinsic disorder, as well as interactions in graphene seems to be essential in understanding the experiments and also in designing graphene-based electronic devices. \\
The failure of the random-phase approximation \cite{Mishchenko}, leads one to employ some non-perturbative methods to investigate the effect of particle-particle interaction and disorder in graphene. An important observation that needs such a method is the appearance of strong carrier density inhomogeneity with density fluctuations much larger than the average density of the system for low densities \cite{HwangAdamSarma}, i.e. Electron-hole puddles (EHPs). EHPs were theoretically predicted by Hwang \textit{et al} \cite{HwangAdamSarma} and Adam \textit{et al} \cite{Adam} as the phase of low carrier density. The existence of these inhomogeneities, characterized by strong electron density fluctuations, were also confirmed in experiments in the vicinity of the Dirac point \cite{Martin,Rutter,Brar,ZhangBrar,Deshpande1,Martin2,Deshpande2,Ishigami,ChoFuhrer,Berezovsky1,Berezovsky2}. The large density fluctuation in this phase were experimentally shown by Martin \textit{et al.} \cite{Martin}. From the comparison of $dI/dV$ map and topography of a sample an interesting observation was made: the rippling of graphene are independent of the charge density inhomogeneities, i.e. EHPs \cite{ZhangBrar}. It was also observed that the spatial extension of puddles is $\approx 20$ nm \cite{ZhangBrar}, consistent with the micro-scale experiment of Martin \textit{et al.} \cite{Martin}.

The observed EHPs are believed to be responsible for the observed minimum conductivity of graphene for which disorder and particle-particle interaction play role simultaneous. In this case (around the zero gate voltage) the transport is governed by the complex network of small random puddles with semimetal character, depending on the details of the charged impurity configuration in the sample. It has been proposed that such inhomogeneity dominates the graphene physics at low ($\lesssim 10^{12}$ cm$^{-2}$) carrier densities \cite{Rossi} in which self-consistent Thomas-Fermi-Dirac (TFD) theory was employed to simulate the graphene charge profile on the SiO$_2$ substrate. The ultimate limit (zero chemical potential) is expected contain very different physics relative to high-density limit, since the charge fluctuation is maximal in this limit, which is not understood properly yet. \\
One important question in the graphene physics is the \textit{existence or absence of the carrier charge self-similarity} which is expected to present in scale-free systems \cite{Najafi1,Najafi2,Najafi3,Najafi4}. Graphene as a zero-gap system has the chance to carry this property in the zero chemical potential. In characterizing this \textit{random surface} the interaction and the disorder should be treated on an equal footing. In the zero gap, the graphene sheet may be viewed as the scale invariant random surface with some scaling relations. The characterization of these surfaces is via determining various exponents and distribution functions.

There are increasing numerical and experimental evidences that many physical phenomena often show scaling relations from the statistical point of view \cite{sornette,falconer,surface2}. Identifying scale invariance symmetry is one of the most important problem of the statistical physics of fluctuating systems, i.e. rough surfaces and surface growth processes \cite{kondevprl,kondevpre,kondevother}, and many other random fluctuating systems. Recently, it was suggested that iso-height lines in these types of random fluctuating fields in $(2+1)-$ dimensions are scale invariant and their size distribution is characterized by a few scaling functions and scaling exponents \cite{kondevpre}. What we are going to do in this paper is confirming this idea for the contour lines of the electron-hole density in Graphene. We will show numerically that within TFD theory, zero-gated graphene is marginally self-similar random surface and have peculiar scaling properties, satisfying hyper-scaling relations \cite{kondevpre}. Recently an attempt concerning this point was made in which it was claimed that the contour lines in graphene membranes are also conformally invariant \cite{herman2016}. \\

The paper is organized as follows. In the next section we will introduce the model and obtain the probability measure of carrier density in weak coupling limit. In the third section we will fix the notation and introduce different scaling behaviors and scaling exponents corresponding to the contour loop ensembles. In the fourth section we will numerically measure the proposed scaling exponents for the disorder potential and the carrier density in Graphene, and we will check the universality of those relations. 
In the final section, we summarize the obtained results and our conclusions.
\section{Ground state of Graphene}
The experimental observation of EHPs is the base of many density-based theories searching for the electronic properties of graphene which is the main motivation of the present paper. Besides the subtleties concerning experimental characterization of mono-layer graphene (MLG) \cite{Rutter,Martin,ZhangBrar,Deshpande1,Martin2,Brar2} and bi-layer graphene (BLG) \cite{Deshpande2}, the coexistence of interaction and (in-plane and out-plane) disorder makes this system less tractable theoretically \cite{SarmaRevModPhys}. Many theoretical attempts have been made to capture the electronic structure of graphene in the presence of disrder, especially the physics of EHPs \cite{Polini,HwangAdamSarma,Rossi}, each of which has its own strengths and weaknesses, for review see \cite{SarmaRevModPhys}. Despite this theoretical background, an overall characterization of the EHPs, especially at the scale-invariant (zero-gate) lavel is missing yet. In this section we present the experimental and theoretical background of EHPs.
\subsection{Experiments}
\begin{figure}[t]
\begin{center}
\includegraphics[width=70mm]{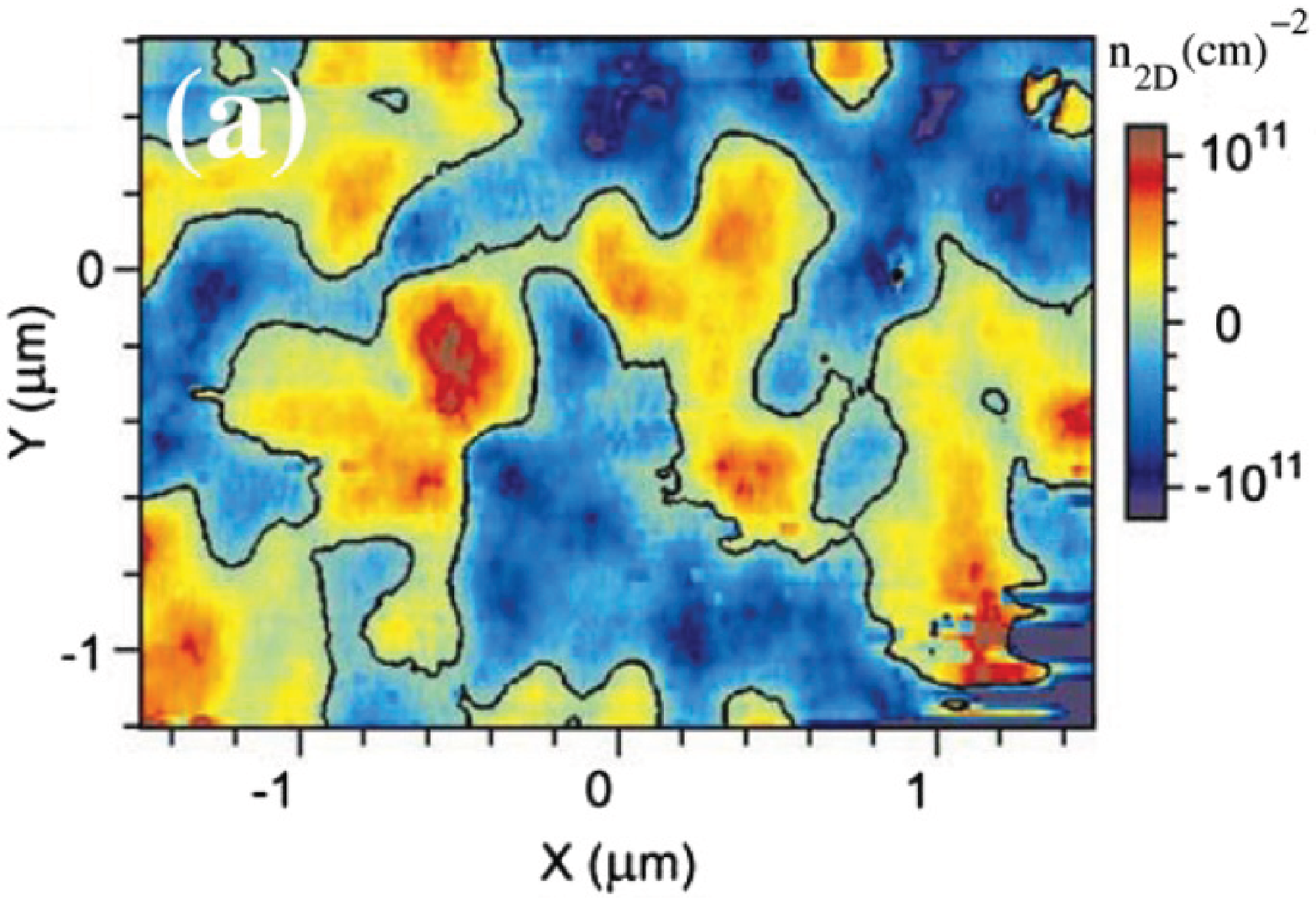}
\includegraphics[width=50mm]{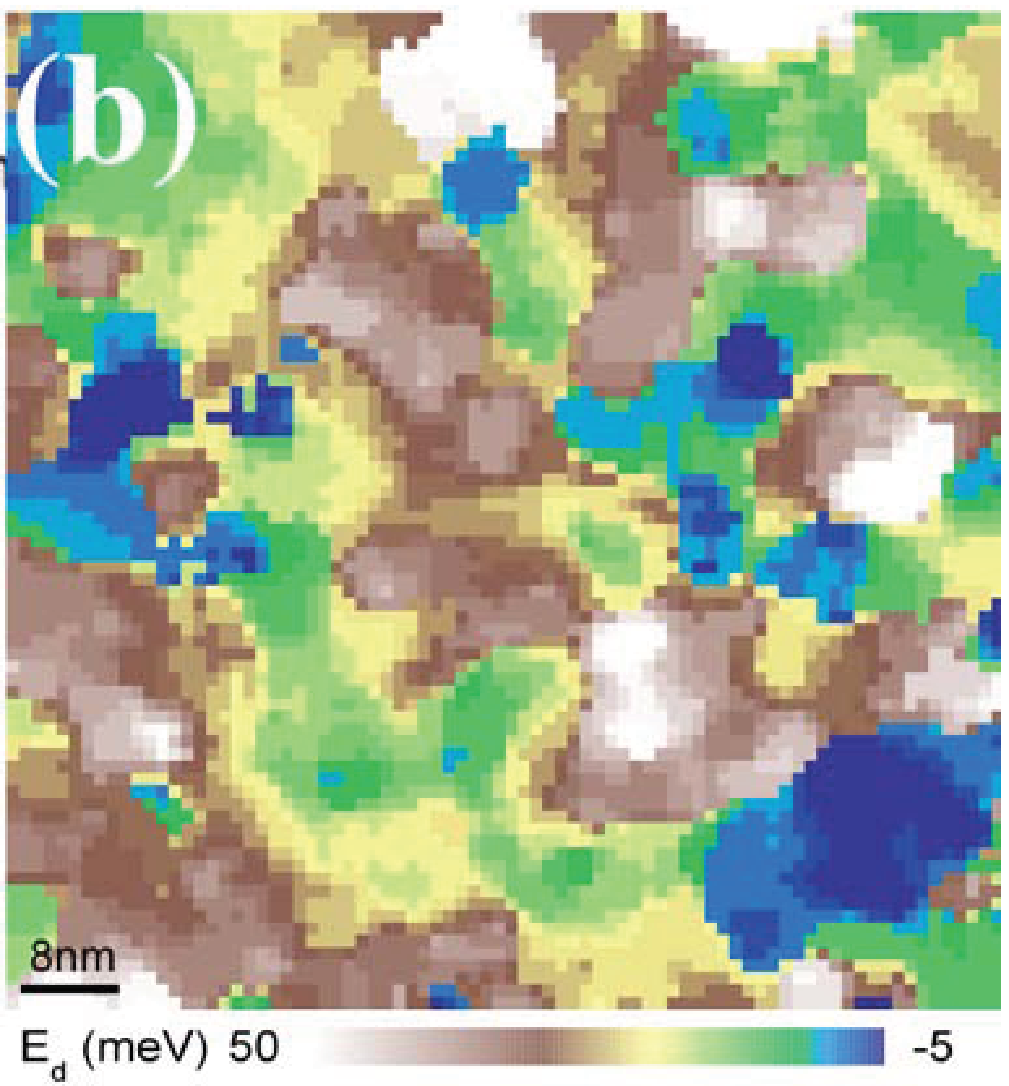}
\end{center}
\caption{(Color online) (a) Carrier density profile at the Dirac point measured with an SET. Adapted from Martin \textit{et al} \cite{Martin}. (b) A 80 nm sample of energy shift of the Dirac point in BLG from STM map. Adapted from Deshpande \textit{et al} \cite{Deshpande2}}
\label{samples}
\end{figure}
In graphene the carrier density is controlled by the gate voltage $n=\kappa_SV_g/4\pi t$ in which $\kappa_S$ is the substrate dielectric constant and $t$ is its thickness and $V_g$ is the gate voltage. The experimental data show a strong dependence on $x\equiv n/n_i$ in which $n_i$ is the impurity density. In ordinary densities, the conductivity is linear function of $x$ and for very low $x$'s, it reaches a minimum of order $\sigma\sim e^2/h$ which is linked to the formation of EHP's. The first scanning probe experiment on exfoliated graphene on SiO$_2$ substrate were done by Ishigami \textit{et al} \cite{Ishigami} revealing its atomic structure and nano-scale morphology. Martin \textit{et al} used the scanning single-electron transistor (SET) to investigate the atomic structure and charge profile of exfoliated graphene close to the Dirac point. Interestingly a high electron density inhomogeneity, breaking up the density landscape in electron-hole puddles were observed in this experiment supporting the theoretical predictions of Adam \textit{et al} \cite{Adam} and Hwang \textit{et al} characterized by large scale electron density fluctuations \cite{HwangAdamSarma}. This strong fluctuations bring the system into a new phase with broken homogeneity in which some random electron and hole conducting puddles are created \cite{HwangAdamSarma}. In the Fig. \ref{samples}a a sample has been shown. The contour lines separate positive $n$ regions from negative ones. These interfaces contain some valuable information about the system in hand. Some attempts for theoretical description of this phenomena was made afterwards \cite{Polini,Rossi}. The low-density minimum conductivity of graphene is believed to be related to the presence of EHPs \cite{Rossi}.\\
A substantial feature of Fig. \ref{samples}a is the formation of large (spanning) clusters of negative or positive charge densities. This feature is also seen in the other (albeit more clean) synthesized samples. A BLG sample in nano-scale, near the Dirac point, has been shown in Fig. \ref{samples}b \cite{Deshpande2} in which the spanning clusters is evident. The presence of the spanning cluster in a system may be the fingerprint of a subtle symmetry; the scale invariance. If true, the system in hand lies within some universality class of the critical phenomena. Such a characterization has not been done for graphene.\\
By analyzing the width in density of the incompressible bands in the quantum Hall regime and fitting the broadened incompressible bands with Gaussian distribution, Martin \textit{et al.} extracted the value of the amplitude of the density fluctuations to be $2.3\times 10^{11}\text{cm}^{-2}$. By calculating the density fluctuations in two ways, namely the probability distribution of the density extracted from the imaging results and the broadening of the incompressible bands in the quantum Hall regime, Martin \textit{et al.} found that the upper bound for the characteristic length of the density fluctuations is $30$ nm, consistent with the theoretical results \cite{Rossi}. Note that in this regime the density fluctuations are much larger than the mean electron density, signaling a different phase from the homogeneous (Dirac) electron gas. In this phase the translational symmetry of the system is broken by forming puddles of electrons and holes, i.e. EHPs.\\
EHPs has been observed and analyzed further using other (direct and indirect) techniques \cite{ChoFuhrer,ZhangBrar,Martin2,Berezovsky1,Berezovsky2}. The first STM experiments on exfoliated graphene showed that in current exfoliated graphene samples the rippling of graphene are independent of the charge density inhomogeneities, i.e. EHPs \cite{ZhangBrar}. This is directly observed from the $dI/dV$ map and topography of a sample which are independent. It was also observed that the spatial extension of puddles is $\approx 20$ nm. The relation between local curvature of the MLG $h(r)$ and the local shift in the Fermi energy ($\delta E_F$) is expressed via $\delta E_F=-\alpha \frac{3a^2}{4}\left(\nabla^2 h(r)\right)^2$ in which $\alpha$ is an energy scale equal to $9.23$ eV and $a$ is the lattice constant. Comparing this with the results of $dI/dV$ map, Deshpande \textit{et al.} showed once again this independence. The more quality investigations show the same results \cite{Martin2,Berezovsky1,Berezovsky2}.
\subsection{Thomas-Fermi-Dirac Theory}
Treating simultaneously particle-particle interaction and disorder on an equal footing is a challenging problem in each condensed matter system. The marginal character of particle-particle interaction in graphene has been firstly shown by Gonzalez \textit{et al.} \cite{Gonzalez} according to which the Fermi velocity is logarithmically enhanced. This \textit{exchange-driven} Dirac-point logarithmic singularity in the Fermi velocity in the intrinsic graphene is shown to disappear in the extrinsic case \cite{HwangSarma}. This logarithmic enhancement of Fermi velocity leads the specific heat to be logarithmically suppressed relative to its non-interacting counterpart \cite{Vafek}. There are many other evidences showing that the band chirality in graphene changes substantially the role of particle-particle interaction with respect to the usual 2D electron gas and the vital (unusual) role of the exchange and correlation energies. The enhancement of screening by means of the exchange and correlation in graphene \cite{Polini}, in contrast to the usual cases, is an example. The other example is the suppression of spin and charge susceptibilities which is attributed to the enhancement of net chirality due to Coulomb interactions in lightly doped graphene \cite{Barlas}, as well as the enhancement of screening effect \cite{Gonzalez}. The opposite dependence of exchange-correlation energy to the charge density with respect to parabolic band 2D electron gas is also a source of many differences of graphene from the other systems. While the later favors inhomogeneous densities, the former increases the energy cost of density increases, favouring more homogeneous densities and enhancing screening.\\
The source of disorder and its relevance in the electronic structure of graphene is also an important question to be addressed. The approximately linear dependence of conductivity on carrier density in graphene sheets \cite{Nomura,HwangAdamSarma} indicates that the remote Coulomb impurities are dominant disorder source in most graphene samples. The experimental observation that the spatial pattern of EHPs is not correlated with the topography of the graphene sheets (described in the previous subsection) is another evidence that the remote charges are the dominant disorder source \cite{Barlas}. The inclusion of Coulomb disorder in graphene in the absence of particle-particle interaction were studied by Fogler \textit{et al.} to investigate diffusive and ballistic transport in graphene $p-n$ junction \cite{Fogler}. The disorder in addition to being the main sources of scattering has an additional effect; it locally shifts the Dirac point. It means that even at the zero gate voltage, the Fermi energy is moved to positive or negative values with respect to the charge neutrality (Dirac) point. The other sources of scattering are ripples \cite{NetoKim} and point defects (which is responsible for high-density saturation of conductivity \cite{HwangAdamSarma}) which are not considered in this paper.\\
The case of relevance is an slow (spatial) varying charge density system. An approach similar in sprit to the LDA-DFT is the Thomas-Fermi-Dirac theory which is valid only for the case $|\nabla_r n(\textbf{r})/n(\textbf{r})|\ll k_F(\textbf{r})$ in which $k_F(\textbf{r})$ is the Fermi wave number at position $\textbf{r}$. It has been shown that for the clean graphene in the low density regime $n\rightarrow 0$ the exchange potential goes to zero such as $V_x(n\rightarrow0)\varpropto -\text{sgn}(n)\sqrt{n}\ln |n|$ as well as the correlation potential, for which the proportionality constant will be introduced below \cite{Polini}. Using local density approximation one can prove that the total energy of the graphene for a disorder configuration and a density profile is \cite{SarmaRevModPhys}:
\begin{eqnarray}
\begin{split}
E=&\hbar v_F[\frac{2\sqrt{\pi}}{3}\int d^2r\text{sgn}(n)|n|^{\frac{3}{2}}\\
&+\frac{r_s}{2}\int d^2r\int d^2r^{\prime}\frac{n(\textbf{r})n(\textbf{r}^{\prime})}{|\textbf{r}-\textbf{r}^{\prime}|}\\
&+r_s\int d^2rV_{xc}[n(\textbf{r})]n(\textbf{r})+r_s\int d^2rV_D(\textbf{r})n(\textbf{r})\\
&-\frac{\mu}{\hbar v_F}\int d^2rn(\textbf{r})]
\end{split}
\end{eqnarray}
in which $v_F$ is the Fermi velocity, $r_s\equiv e^2/\hbar v_F\kappa_S$ is the dimensionless interaction coupling constant, $\mu$ is the chemical potential, $g=g_sg_v=4$ is the total spin and valley degeneracy. The exchange-correlation potential is calculated to be \cite{Polini}:
\begin{eqnarray}
V_{xc}=\frac{1}{4}\left[ 1-gr_s\zeta(gr_s)\right]\text{sgn}(n)\sqrt{\pi|n|}\ln\left(4k_c/\sqrt{4\pi |n|}\right)
\end{eqnarray}
in which $k_c$ is the momentum cut-off and $\zeta(y)=\frac{1}{2}\int_0^{\infty}\frac{dx}{(1+x^2)^2\left( \sqrt{1+x^2}+\pi y/8\right)}$. The remote Coulomb disorder potentail is calculated by the relation:
\begin{eqnarray}
V_D(r)=\int d^2r^{\prime}\frac{\rho(\textbf{r}^{\prime})}{\sqrt{|\textbf{r}-\textbf{r}^{\prime}|^2+d^2}}
\label{VDis}
\end{eqnarray}
in which $\rho(r)$ is the charged impurity density and $d$ is the distance between substrate and the graphene sheet. For the graphene on the SiO$_2$ substrate, $\kappa_S\simeq 2.5$, so that $r_s\simeq0.8$, $d\simeq 1$ nm, $k_c=1/a_0$ where $a_0$ is the graphene lattice constant $a_0\simeq 0.246$ nm corresponding to energy cut-off $E_c\simeq 3$ eV. It is notable that in the above equations we have considered bare coulomb interactions for both impurity and Hartree terms. This is due to the absence of screening in low career densities, i.e. in the vicinity of the Dirac points. To obtain the equation governing $n(r)$ one can readily minimize the energy with respect to $n(r)$:
\begin{eqnarray}
\begin{split}
&\text{sgn}(n)\sqrt{|\pi n|}+\frac{r_s}{2}\int d^2 \textbf{r}^{\prime}\frac{n(\textbf{r}^{\prime})}{|\textbf{r}-\textbf{r}^{\prime}|}\\
&+r_sV_{xc}[n]+r_sV_D(\textbf{r})-\frac{\mu}{\hbar v_F}=0
\end{split}
\label{mainEQ}
\end{eqnarray}
which should be solved self-consistently. In this paper we consider the disorder to be white noise with Gaussian distribution $\left\langle \rho(\textbf{r})\right\rangle=0 $ and $\left\langle \rho(\textbf{r})\rho(\textbf{r}^{\prime})\right\rangle=(n_id)^2\delta^2(\textbf{r}-\textbf{r}^{\prime})$. Due to pure $1/r$ dependence of the Hartree and disorder terms, the convergence of the equation is slow.
\subsection{Scaling and the Probability Measure}
Let us now concentrate on the scaling properties of this equation excluding $V_{xc}$. By zooming out the system, i.e. the transformation $\textbf{r}\rightarrow \lambda \textbf{r}$, we see that for the case $\mu=0$ the equation remains unchanged if we transform $n(\textbf{r})\rightarrow n(\lambda\textbf{r})=\lambda^{-2}n(\textbf{r})$ as expected from the spatial dimension of $n(\textbf{r})$. This is because of the fact that $V_D(\lambda\textbf{r})=\lambda^{-1}V_D(\textbf{r})$. This symmetry is very important, since it causes the system to be self-affine and may be violated for other choices of disorder. This scale-invariance in two dimensions leads to power-law behaviors and some exponents which are vital for surface characterization. It may also lead to conformal invariance of the system, and if independent of type of disorder, bring the graphene surface into a member of the minimal conformal series. The existence of $V_{xc}$ makes things difficult, since $V_{xc}(\textbf{r})\rightarrow V_{xc}(\lambda\textbf{r})=\lambda^{-1}\left(V_{xc}-\beta\text{sgn}(n)\sqrt{\pi|n|}\ln\lambda\right) $ in which $\beta\equiv\frac{1}{4}(1-gr_s\zeta(gr_s))$. Therefore the rescaled equation is:
\begin{eqnarray}
\begin{split}
&\xi(\lambda)\ \text{sgn}(n)\sqrt{|\pi n|}+\frac{r_s}{2}\int d^2 \textbf{r}^{\prime}\frac{n(\textbf{r}^{\prime})}{|\textbf{r}-\textbf{r}^{\prime}|}\\
&+r_sV_{xc}[n]+r_sV_D(\textbf{r})=0
\end{split}
\end{eqnarray}
in which $\xi(\lambda)\equiv 1-\beta r_s\ln\lambda$. Therefore the first term survive marginally in the infra-red limit and scale invariance is expected, even in the presence of $V_{xc}$. The above symmetry is simply an additional symmetry which limits the correlation functions to show power-law behaviors, but further details of the system needs exact or numerical solution. One of the most important quantities in random field analysis is the probability measure of charge density $P(n)$. It is believed that the probability measure of charge density in graphene is not Guassian \cite{SarmaRevModPhys}. In the remaining of this subsection we search for analytical form of $P(n)$ (or $P_n$) and present the result for the case of very weak coupling limit in some approximation.\\
Let us now search for the analytic for of probability measure of density $P_n$ by focusing on Eq. \ref{mainEQ}. By analytic continuation of $n$ to complex variables, and noting that:
\begin{eqnarray}
\begin{split}
& \nabla(\text{sgn}(n)\sqrt{\pi |n|})=\frac{1}{2}\text{sgn}(n)\sqrt{\frac{\pi}{|n|}}\nabla n\\
& \nabla V_{xc}=-\frac{1}{2}\beta\sqrt{\frac{\pi}{|n|}}\left[ 1-\text{sgn}(n)\ln\left( \frac{4k_c}{\sqrt{4\pi |n|}}\right) \right] \nabla n
\end{split}
\label{mainEQ1}
\end{eqnarray}
and using Eq. \ref{mainEQ} we obtain:
\begin{eqnarray}
\begin{split}
& \nabla n=-r_sF_n\left\lbrace \overrightarrow{\chi}_{\rho}(d)+\frac{1}{2}\overrightarrow{\chi}_n(0)\right\rbrace 
\end{split}
\end{eqnarray}
in which $F_n\equiv\frac{2\text{sgn}(n)\sqrt{|n|/\pi}}{1-r_s\beta\left[\text{sgn}(n)-\ln\frac{4k_c}{\sqrt{4\pi |n|}}\right]}$ and $\overrightarrow{\chi}_x(d)\equiv \int d^2\textbf{r}^{\prime}x(\textbf{r}^{\prime})\nabla \left(|\textbf{r}-\textbf{r}^{\prime}|^2+d^2\right)^{-1/2}$ and $x=\rho , n$.
Now we perform some Ito calculations to obtain the probability measure of $n(r)$. The differential of the charge density is obtained via ($\text{d}\chi_x(d)\equiv\int d^2\textbf{r}^{\prime}\text{d}\left[x(\textbf{r}^{\prime})\right] \left(|\textbf{r}-\textbf{r}^{\prime}|^2+d^2\right)^{-1/2}$):
\begin{eqnarray}
\begin{split}
& \text{d}n=-r_sF_n\left[\text{d}\chi_{\rho}(d)+\frac{1}{2}\text{d}\chi_n(0)\right] 
\end{split}
\end{eqnarray}
(In this formula and the remaining, the integral differential and the external differential variables are shown respectively by lower case, e.g. $dx$ and upper case, e.g. $\text{d}x$). Let us suppose that $f$ is an arbitrary local or non-local smooth function of the density $n$, i.e. $f_{\textbf{r}_0}[n]=\int d^2\textbf{r}f_{\textbf{r}_0}(n(\textbf{r}))g(\textbf{r},\textbf{r}_0)$ in which $\textbf{r}_0$ is the origin from which $\textbf{r}$ is measured and $g(\textbf{r},\textbf{r}_0)$ is some (arbitrary) function of $|\textbf{r}-\textbf{r}_0|$. Without loose of generality we set $g\equiv1$ to facilate the calculations. $f$ can therefore be readily expanded in terms of $n$ using the the above equations: 
\begin{eqnarray}
\begin{split}
&\text{d}f_{\textbf{r}_0}[n]\equiv \int d^2\textbf{r}[f_{\textbf{r}_0}(n(\textbf{r})+\text{d}n(\textbf{r}))-f_{\textbf{r}_0}(n(\textbf{r}))]\\
&=\int d^2\textbf{r}[\partial_nf_{\textbf{r}_0}(n(\textbf{r}))\text{d}n(\textbf{r})+\frac{1}{2}\partial_n^2 f_{\textbf{r}_0}(n(\textbf{r}))\text{d}n(\textbf{r})^2]\\
& =\int d^2\textbf{r}\left[ -r_sF_n\left( \text{d}\chi_{\rho}(d)+\frac{1}{2}\text{d}\chi_n(0)\right) \right]\partial_n f_{\textbf{r}_0}(n(\textbf{r}))\\
&+\int d^2\textbf{r}\left[ \frac{r_s^2}{2}F_n^2(\text{d}\chi_{\rho})^2\right]\partial_n^2 f_{\textbf{r}_0}(n(\textbf{r}))
\end{split}
\end{eqnarray}
When $\text{d}n$ is purely due to spatial changes, then we have $\text{d}f_{\textbf{r}_0}[n]=\int d^2\textbf{r}[f_{\textbf{r}_0+d\textbf{r}}(n(\textbf{r}))-f_{\textbf{r}_0}(n(\textbf{r}))]$, i.e. one changes the view point from active to passive. We can calculate the probability measure of the density, noting that the average value of $f$ should not depend on $\textbf{r}_0$ due to homogeneity of the system. The change of the average, due to changing the origin is ($\left\langle f\right\rangle \equiv\int \prod_{\textbf{r}^{\prime}}dn(\textbf{r}^{\prime})P_{\textbf{r}_0}(\left\lbrace n\right\rbrace)f(n(\textbf{r}^{\prime}))$):
\begin{eqnarray}
\begin{split}
\text{d}\left\langle f\right\rangle &=\int d^2\textbf{r}\left\langle \left[ -r_sF_n(d\chi_{\rho}+\frac{1}{2}d\chi_n)\right] \partial_n f \right. \\
&\left. +\frac{r_s^2}{2}F_n^2(d\chi_{\rho})^2\partial_n^2f \right\rangle \\
\end{split}
\end{eqnarray}
On the other hand, one can write the above equation in terms of $P_{\textbf{r}_0}(\left\lbrace n\right\rbrace)$ as $\text{d}\left\langle f\right\rangle =\int \prod_{\textbf{r}^{\prime}}dn(\textbf{r}^{\prime})\text{d}P_{\textbf{r}_0}(\left\lbrace n\right\rbrace)f(n(\textbf{r}^{\prime}))$. Using this fact and noting that $\left\langle d\chi_{\rho}=0\right\rangle$, and representing the averages as the integrals over probability measures, and doing integration by parts, one finds:
\begin{eqnarray}
\begin{split}
\text{d}P_{\textbf{r}_0}(\left\lbrace n\right\rbrace )=\frac{1}{2}r_s\partial_n\left[ F_n\text{d}\chi_nP_n\right] +\frac{r_s^2}{2}\partial_n^2\left[F_n^2(\text{d}\chi_{\rho})^2P_n\right] 
\end{split}
\end{eqnarray}
Therefore for the homogenous system, which is independent of the choice of the observation point $\textbf{r}_0$, the equation governing the distribution of $n(r)$, considering particle-hole symmetry, is:
\begin{eqnarray}
\begin{split}
r_s\partial_n\left[F_n^2(\text{d}\chi_{\rho})^2P_n\right] =- F_n\text{d}\chi_nP_n
\end{split}
\end{eqnarray}
Now if we calculate $\text{d}\chi_n(0)\equiv \nabla \chi_n(0).\text{d}\textbf{r}$ and $(\text{d}\chi_{\rho}(d))^2\equiv (\nabla \chi_{\rho}(d).\text{d}\textbf{r})^2$ and replace $\cos^2\theta$ by $\frac{1}{2}$, we see that to first order of $\text{d}r$, $\left\langle (\text{d}\chi_{\rho})^2\right\rangle =\frac{\pi dn_i^2}{2\sqrt{2}}\text{d}r$ and $\text{d}\chi_n(0)=\frac{1}{\sqrt{2}}G_n \text{d}r$ in which $G_n=\int d^2\textbf{r}^{\prime}\frac{n(\textbf{r}^{\prime})}{|\textbf{r}-\textbf{r}^{\prime}|^2}$. Finally we obtain
\begin{eqnarray}
\partial_nP(\left\lbrace n\right\rbrace ) =- \Gamma_n P(\left\lbrace n\right\rbrace )
\label{MainPnEq}
\end{eqnarray}
in which $\zeta\equiv \frac{2}{\pi dn_i^2r_s}$ and $\Gamma_n\equiv\zeta\frac{G_n}{F_n}+2\partial_n\ln F_n $ and $\partial_n$ is the functional derivative. This equation is the master equation governing the probability distribution of a density configuration which should clearly contains derivatives of the charge density. For the local charge probability distribution $P(n)$, the calculations is much simpler than above. For this case it is sufficient to carry out Ito calculations on some local function of charge density, i.e. $f(n)$ and use the independence of $P(n)$ of the spatial point $\textbf{r}$. The result is the same as Eq. \ref{MainPnEq}, replacing $P_n(\left\lbrace n\right\rbrace )$ simply by $P(n)$ and the functional derivative by simple derivative, i.e. $\partial_nP(n) =- \Gamma_n P(n)$.\\
One may be interested in the solution of the above equation for weak coupling limit $r_s\rightarrow 0$, or the weak disorder limit $n_i\rightarrow0$, i.e. large $\zeta$ limit. In this limit, and considering $G_n$ to be nearly constant, i.e. $G_n\sim \left\langle n\right\rangle$, we have $\Gamma_n\simeq \frac{\sqrt{\pi}}{2}\zeta^{\prime}\frac{\text{sgn}(n)}{\sqrt{|n|}}=\zeta^{\prime}\partial_n\left(\text{sgn}(n)\sqrt{\pi|n|}\right)$ in which $\zeta^{\prime}\equiv \zeta G_n$. The solution is therefore:
\begin{eqnarray}
\begin{split}
P_n=A\exp\left[-\zeta^{\prime}\left(\text{sgn}(n)\sqrt{\pi |n|}-\frac{\mu}{\hbar Sv_F}\right)\right]
\end{split}
\label{PnZeroth}
\end{eqnarray}
in which $A$ is a normalization constant and $S$ is the area of the sample. This relation may seem to be unsuited, since it grows unboundedly for negative values of $n$. Actually there is no contradiction, due to the presence of $G_n$ whose amount grows negatively for negative $n$ values, which returns the above equation into expected form. In fact the original charge equation has electron-hole symmetry for the case $\mu=0$ which should result to an electron-hole symmetric form of $P_n$. Our approximation (considering $G_n$ as a constant) violated this symmetry. Re-considering this quantity as a dynamical variable retains the mentioned symmetry. It is also notable that the second term in the exponent ($\frac{\mu}{\hbar v_F}$) has been inserted due to some symmetry considerations and the above equation satisfies the original equation of $P_n$.\\
In the above equation, the effects of disorder and Hartree interaction have been coded in $\zeta^{\prime}$. It is clear that a very weak interaction, has the same effect as a very weak disorder, i.e. in both cases ${\zeta^{\prime}}^{-1}\rightarrow \infty$ which results to very wide charge distribution and large charge fluctuations. The other limit which is our main concern is the $\mu\rightarrow 0$ limit which has direct effect on $G_n$. In fact $\mu$ controls $\left\langle n\right\rangle$ which directly affects $\int d^2\textbf{r}^{\prime}\ \frac{n(\textbf{r}^{\prime})}{|\textbf{r}-\textbf{r}^{\prime}|}$, i.e. $G_n$. In the limit $\mu\rightarrow 0$, one expects that $G_n$ becomes vanishingly small, so that ${\zeta^{\prime}}^{-1}\rightarrow \infty$ which implies large scale density fluctuations. This is the point we emphasized in previous sub-sections: at the Dirac point the density fluctuations grow unboundedly which drives the system into a new phase, i.e. formation of EHPs, consistent with other theoretical results \cite{Rossi}. In this limit the power-law behaviors become possible.\\
In the above equation we ignored $V_{xc}$ and considered the limit of small graphene fine structure constant, i.e. $r_s\rightarrow 0$ which gives some sense about the behavior of the probability measure. The inclusion of $V_{xc}$ and extending the analysis to all range of $r_s$ make the Eq. \ref{PnZeroth} invalid, so that $P(n)$ may show different dependence on $n$ for other ranges of $r_s$. The Eq. \ref{MainPnEq} should be solved non-perturbly in this case which is beyond our analysis. To this end, we have solved numerically the Eq. \ref{mainEQ} which is the subject of the following sections. First we introduce the random field rough surfaces in the next section, and then present our results in terms of this framework.
\section{Scaling properties of contour loop ensembles}
Graphene may be viewed as a two-dimensional (2D) random-field media in which the charge profile and also the impurity coulomb potential are viewed as the random fields. We argued that at $\mu=0$ the Eq. \ref{mainEQ} is scale invariant, i.e. $n(\lambda\textbf{r})\stackrel{d}{=}\lambda^{-2}n(\textbf{r})$ in which $\stackrel{d}{=}$ means the equality of the distributions. This may be interpreted as the signature of the scale invariance of our 2D random field. Before we proceed, it seems necessary to review some features of the scale-invariant 2D random rough fields which is the aim of this section. \\
Let $h(x,y)\equiv h(\mathbf{r})$ be the \textit{height} profile (in the graphene case, the charge profile or the impurity coulomb potential) of a scale invariant 2D random rough field. The main property of self-affine random fields is their invariance under rescaling \cite{surface2,falconer,sornette}. In other words the probability distribution function is such that the random profile $h(\mathbf{r})$ has self-affine
scaling law
\begin{eqnarray}
h(\lambda \mathbf{r}) \stackrel{d}{=} \lambda ^{\alpha} h(\mathbf{r}),
\end{eqnarray}
where the parameter $\alpha$ is \textit{roughness} exponent or the \textit{Hurst} exponent and $\lambda$ is a scaling factor. The translational, rotational and scale invariance of $h(\mathbf{r})$ imply that the height-correlation function of random fields behaves as
\begin{eqnarray}\label{height-corr}
C(r) \equiv \langle \left[ h(\mathbf{r}+\mathbf{r_0})-h(\mathbf{r_0}) \right]^2 \rangle \sim |\mathbf{r}| ^{2\alpha_l},
\end{eqnarray}
where the parameter $\alpha_l$ is called the local roughness exponent \cite{surface2} and $\left\langle \right\rangle$ denotes the ensemble average. Another measure to classify the scale invariant profile $h(\mathbf{r})$ is the total variance
\begin{eqnarray}\label{total variance}
W(L)\equiv \langle \left[ h(\mathbf{r}) - \bar{h} \right]^2 \rangle _L \sim L^{2\alpha_g}
\end{eqnarray}
where $\bar{h}=\langle h(\mathbf{r}) \rangle_L$, and $\langle \dots \rangle _L$ means that, the average is taken over $\mathbf{r}$ in a box of size $L$. The parameter $\alpha_g$ is the global roughness exponent. Self-affine surfaces are mono-fractals just if $\alpha_g = \alpha_l = \alpha$ \cite{surface2}.
In general, the scaling properties of the height-correlation function Eq. (\ref{height-corr}) as well as the $Hurst$ exponent $\alpha$, are the most important quantities to distinguish a given mono-fractal random field from the others. The scaling properties of the two point correlation function Eq. (\ref{height-corr}) gives the scaling relation for the Fourier power spectrum, {i.e.}, $S(\mathbf{q})\equiv \langle |h(\mathbf{q})|^2\rangle\sim |\mathbf{q}|^{-2(1+\alpha)}$, for small values of $q$ or large values of $r$ in which $h(\textbf{q})$ is the Fourier transform of $h(r)$ \cite{falconer}. A wide variety of mono-fractal random fields with roughness exponent $\alpha$ are governed by a Gaussian distribution 
\begin{eqnarray}
\mathcal{P}\left[ h \right]\sim \exp \left[ -\frac{k}{2} \int _ 0 ^ {\Lambda} d^2 \mathbf{q}\mathbf{q}^{2(1+\alpha)}h_{\mathbf{q}}h_{-\mathbf{q}} \right],
\end{eqnarray}
where $\Lambda$ is the high momentum cut-off and $h(\mathbf{q})$ is the Fourier transform of $h(\mathbf{r})$ and $k$ is the stiffness.
\begin{figure}[t]
\begin{center}
\includegraphics[width=45mm]{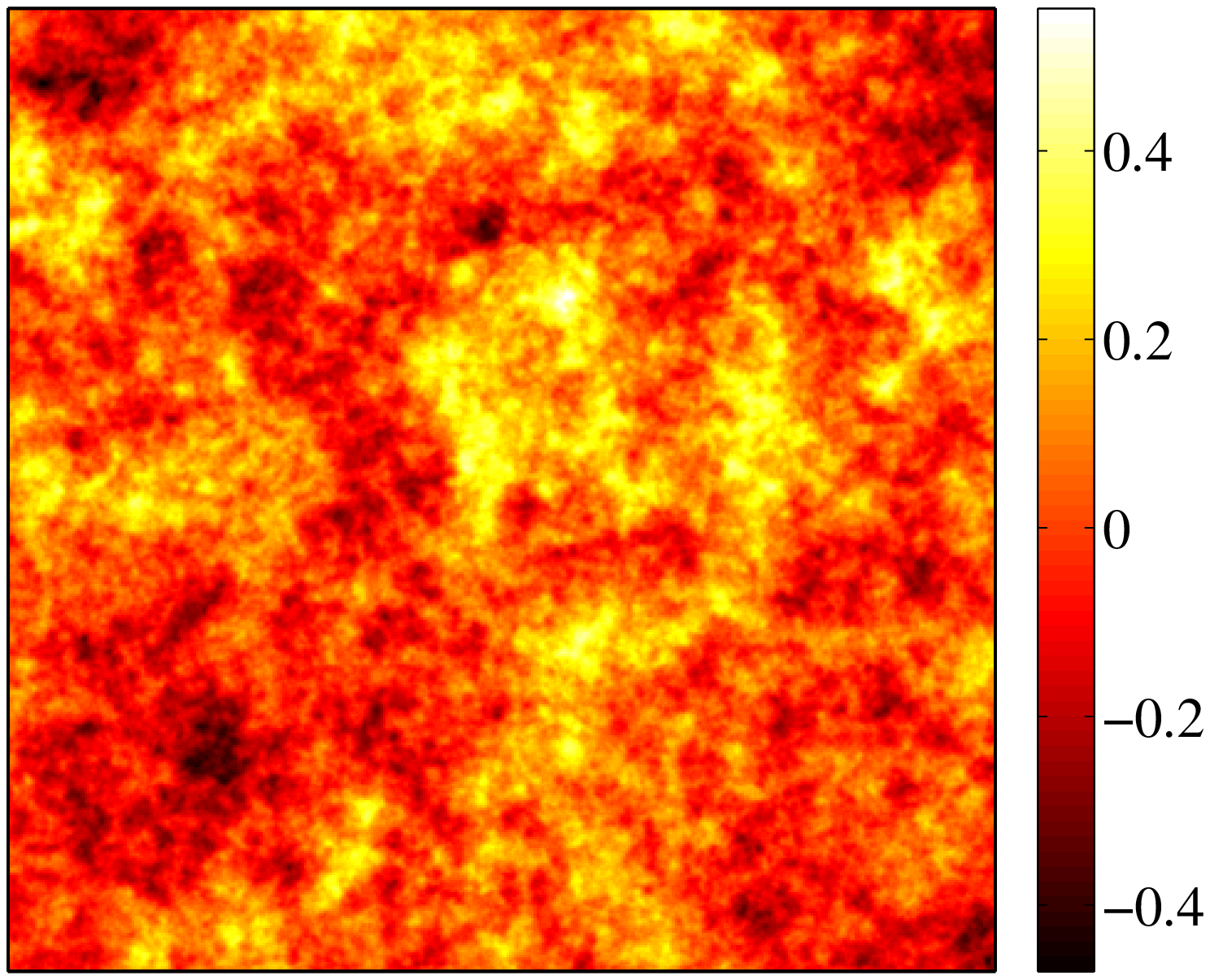}
\includegraphics[width=40mm]{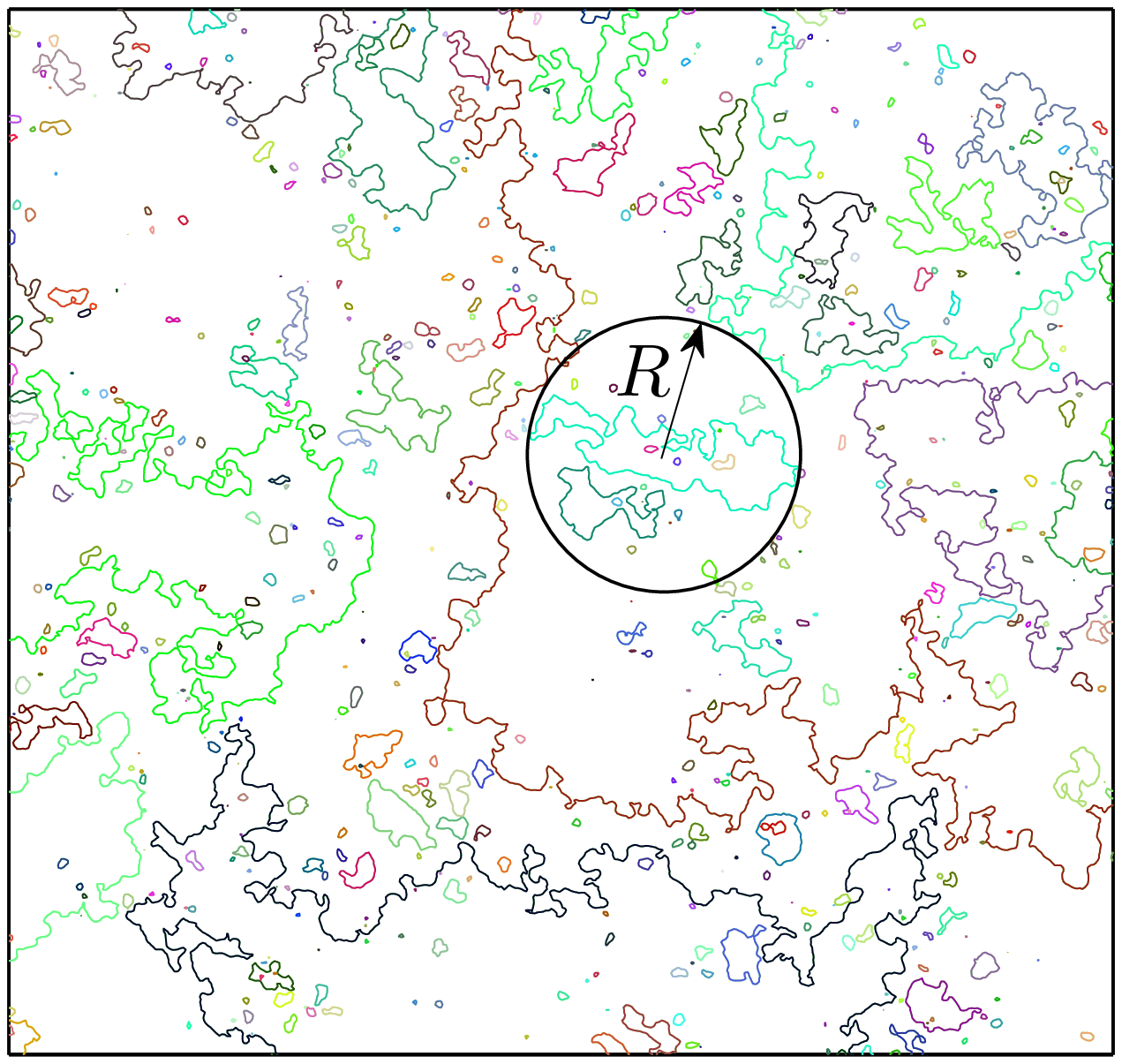}
\\
\includegraphics[width=45mm]{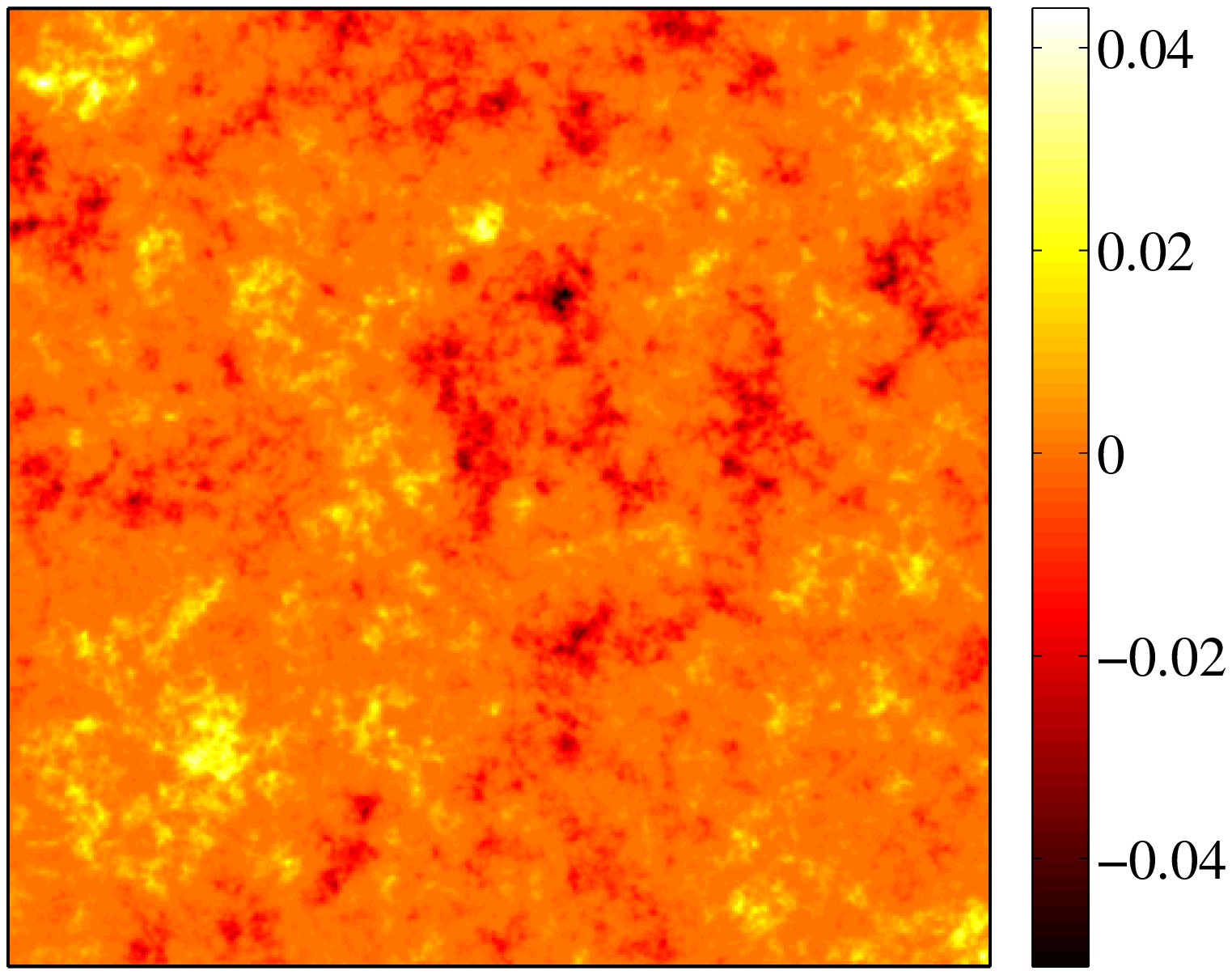}
\includegraphics[width=40mm]{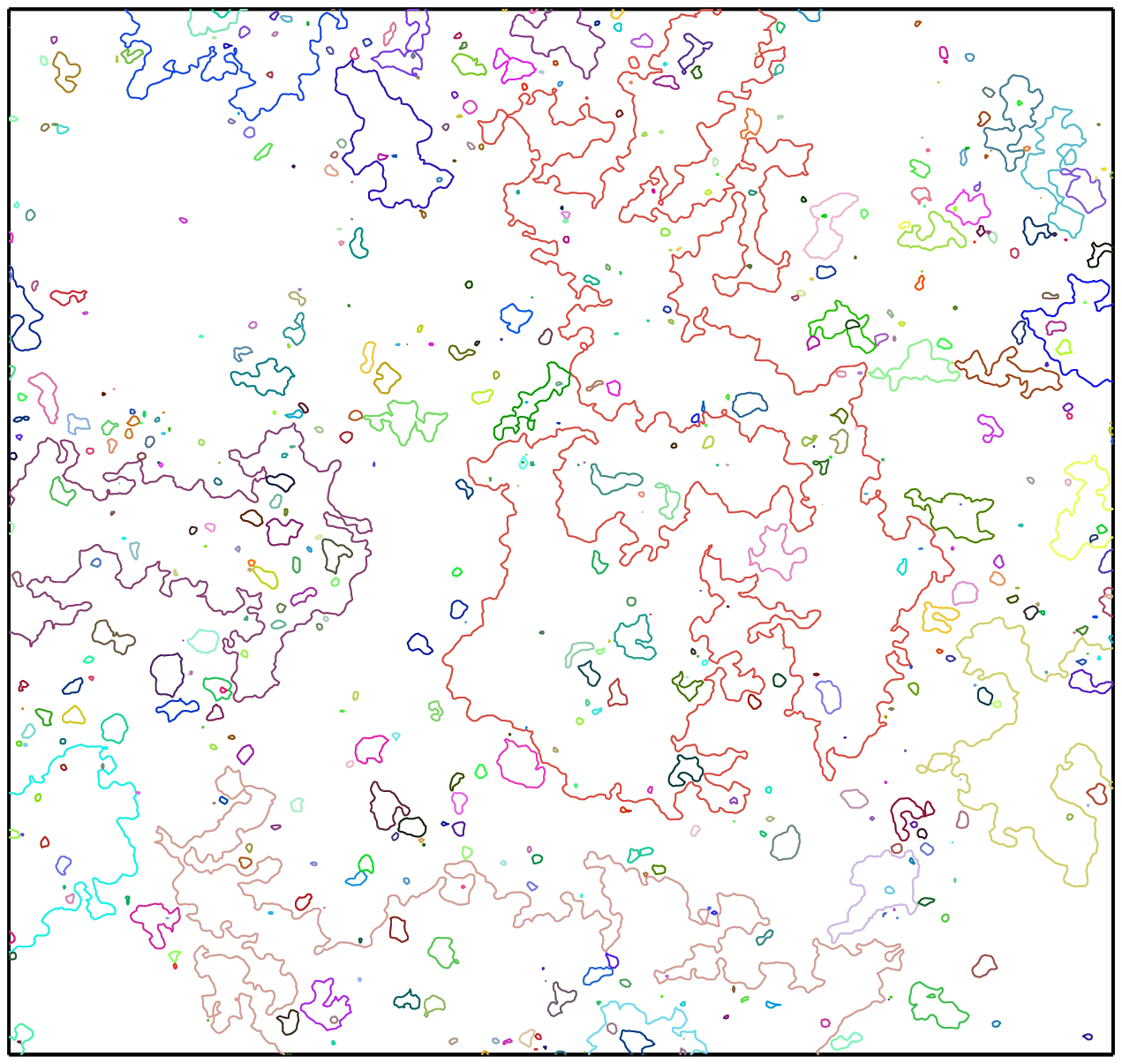}
\end{center}
\caption{(Color online) Color plot and mean-level contour lines of the disorder potential $V_D$ (top). 
Color plot and mean-level contour lines of carrier density distribution $n(x,y)$ for Graphene at the Dirac point (bottom). The contour plot consists of closed non-intersecting loops that
connects points of $\bar{n} = 0$. For $n > 0$ ($n < 0$) we have particles (holes).}
\label{figur1}
\end{figure}
One of the most interesting characteristics of a mono-fractal random Gaussian surface is the scaling properties of the iso-height lines of the rough profile $h(\mathbf{r})$ at the level set $h(\mathbf{r}) = h_0$. The intersection between the self-affine surface $h(x,y)$ and a horizontal plane perpendicular to the $z$ axes, contains many closed non-intersecting loops or a configuration of $contour$ $ensemble$ which come in many shapes and sizes \cite{kondevprl,kondevpre}. These geometrical objects are scale invariant and one can focus to study the non-local features of the contour loops, i.e., 
their size distribution is characterized by a few power law relations and scaling exponents. The scaling theory of contour loop ensembles of self-affine Gaussian fields was introduced in Ref. \cite{kondevprl} and developed in Ref. \cite{kondevpre}. Following Ref. \cite{kondevpre}, here we introduce different scaling laws and scaling exponents. The contour loop ensemble can be characterized through the loop correlation function $G_l(\mathbf{r})$ and the probability distribution of contours $\tilde{n}(s,R)$ in which $s$ is the loop length and $R$ is the loop radius. In fact for every contour loop in the level set, the probability distribution $\tilde{n}(s,R)$ is the measure to have contours with length $(s, s + ds)$ and radius $(R, R + dR)$. The loop correlation function is a probability measure of how likely the two points separated by the distance $r$ lie on the same contour. The loop correlation function is considered to be rotationally invariant that forces $G_l(r)$ to depend only on $|r|$. This probability function for the contour loop ensembles on the lattice with grid size $a$ and in the limit $r \gg a$ scales with $r$ as 
\begin{eqnarray}\label{loop correlation function}
G_l(r) \sim \frac{1}{r^{2x_l}},
\end{eqnarray}
where $x_l$ is the loop correlation exponent. It is believed that the exponent $x_l$ is the superuniversal quantity and for all the known mono-fractal Gaussian random fields in two dimensions this exponent is equal to $x_l=\frac{1}{2}$ \cite{kondevprl,kondevother,haghighi2011,hosseinabadi2012,hosseinabadi2014}. That the contour loop ensemble is scale invariant, forces $\tilde{n}(s,R)$ to scale with $s$ and $R$ as
\begin{eqnarray}
\tilde{n}(s,R) \sim s^{-\tau -1/D_f} f_n(s/R^{D_f}),
\end{eqnarray}
where $f_n(s/R^{D_f})$ is a scaling function and the exponents $D_f$ and $\tau$ are the fractal dimension and the length distribution exponent, respectively. 
For the scale invariant contour lines, one can define the fractal dimension as the exponent in the scaling relation between mean contour length $\langle s \rangle $ and the radius $R$. The relation
between the average loop length and the radius is derived from $\tilde{n}(s, R)$ by integration:
\begin{eqnarray}\label{s-R}
\langle s \rangle \equiv \frac{\int _0^\infty s\tilde{n}(s,R) ds}{\int _0^\infty \tilde{n}(s,R) ds} \sim R^{D_f}.
\end{eqnarray}
Note that integrating $\tilde{n}(s,R)$ over all radii gives the probability distribution of contour lengths $\tilde{P}(s)$ that is a probability measure for the contour loops with length $s$. 
The density of loops with length $s$ follows the power law
\begin{eqnarray}\label{pdf of contour size}
\tilde{P}(s) \equiv \int_0^\infty \tilde{n}(s,R) \sim s^{-\tau}.
\end{eqnarray}
For a self affine random field, the cumulative distribution of the number of contours with area greater than $A$ is another interesting quantity with the scaling property. The cumulative distribution of area $N_{>}(A)$ has the scaling form 
\begin{eqnarray}\label{scaling of cumulative dist}
N_>(A)\sim A^{-\zeta/2},
\end{eqnarray}
where for mono fractal contour lines $\zeta = 2-\alpha$. 
The scaling properties of the contour ensemble justifies the scaling relations between five different scaling exponents $\alpha$, $D_f$, $\tau$, $\zeta$ and $x_l$ which satisfy the relations \cite{kondevpre}
\begin{eqnarray}\label{hyper1}
D_f(\tau -1) &=& \zeta \nonumber \\ 
&=&2-\alpha,
\end{eqnarray}
and
\begin{eqnarray}\label{hyper2}
D_f(\tau-3) = 2x_l -2.
\end{eqnarray}
Following from Eqs. (\ref{hyper1}) and (\ref{hyper2}) one
can find two exponents $D_f=2-x_l-\alpha/2$ and $\tau = 1+(2-\alpha)/(2-x_l-\alpha/2)$ as a function of the
$Hurst$ exponent $\alpha$ and the loop correlation exponent $x_l$.
In the next section we will numerically calculate all mentioned exponents $\alpha$, $D_f$, $\tau$ and $\zeta$ for the disorder potential and the electron-hole distribution in Graphene. We will also check the scaling relations Eqs. (\ref{hyper1}) and (\ref{hyper2}) between different exponents. 
\begin{figure}[t]
\begin{center}
\includegraphics[width=8cm,clip]{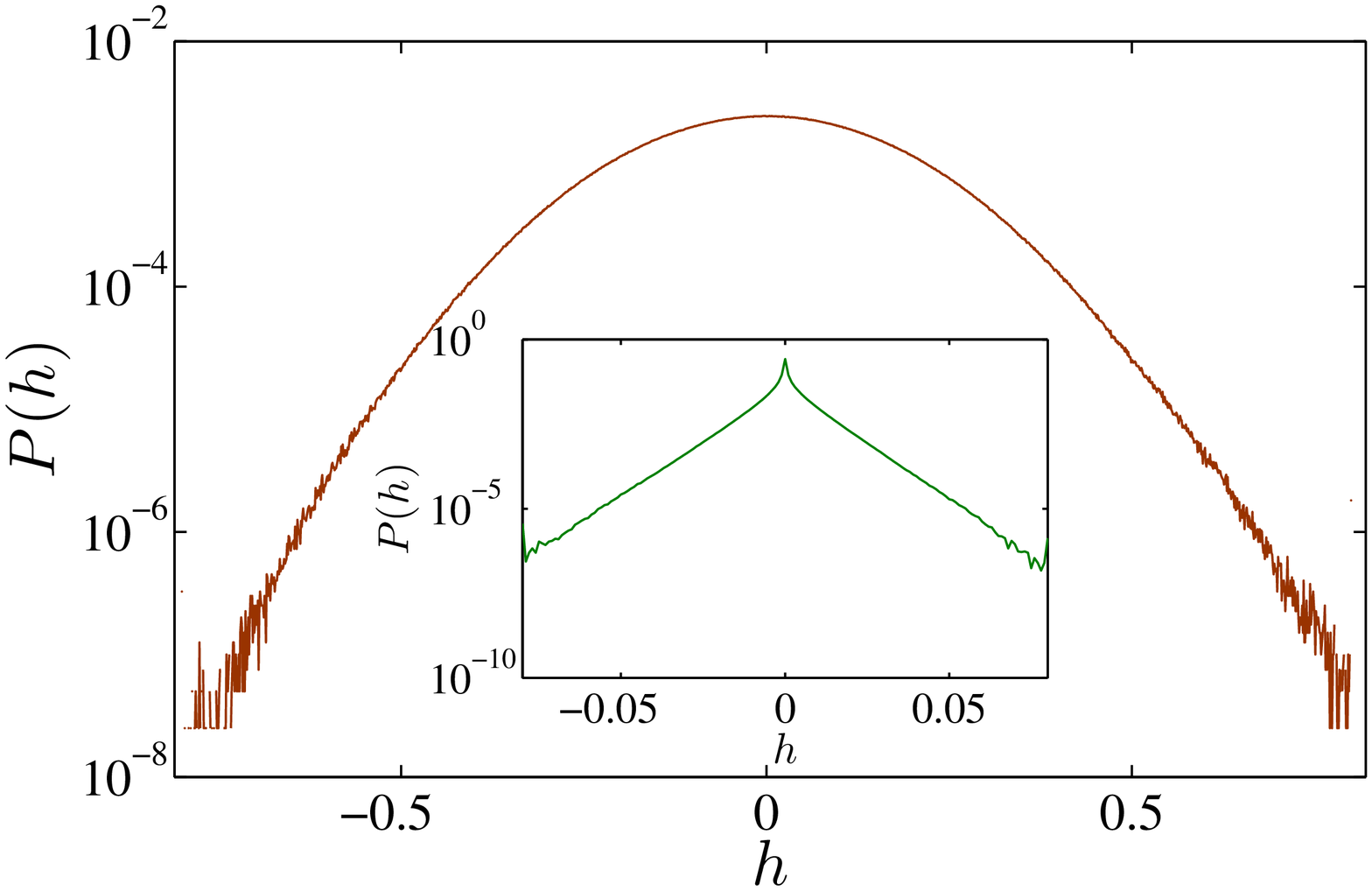}
\end{center}
\caption{(Color online) The Gaussian probability distribution function $\mathcal{P}\lbrace h \rbrace$ of the disorder potential $h=V_D(x,y)$ and non-Gaussian PDF of the carrier density distribution $h=n(x,y)$ for Graphene at the Dirac point (inset) in semi-log scale.}
\label{figur2}
\end{figure}
\section{Numerical results and discussion}
To extract the contour lines of the disorder potential and the corresponding electron-hole distribution at mean level $\bar{h}=\langle h(\mathbf{r}) \rangle_L$ , we use the contouring algorithm followed from \cite{kondevpre}.
In Fig. (\ref{figur1}) we have plotted the mean level contour loop ensembles for $h(\mathbf{r}) = V_D(x,y)$ and the carrier density $h(\mathbf{r}) = n(x,y)$. We would like to measure the scaling exponent $\alpha_l$ and $\alpha_g$ associated with the scaling properties of the height-correlation functions and total variances of the random fields $V_D(\mathbf{r})$ and $n(\mathbf{r})$. Then we will directly measure the scaling exponents $x_l$, $D_f$, $\zeta$ and $\tau$ and we will show that these exponents are universal and depend only to the roughness exponent $\alpha$ of these processes.\\
In our numerical process, we have discretized the real space by $1$ nm steps and generated $L\times L$ square lattice. We have repeated our analysis for $L=50$ nm, $100$ nm, $200$ nm, $300$ nm and $400$ nm to control the finite size effects. We found that the results are independent of the system size for $L\gtrsim 100$ nm. Over $6\times 10^3$ samples for each system size were generated (the total (2.4 GHz) CPU time spent was $1.2\times 10^8$ s). The steepest descent method were used to solve Eq. \ref{mainEQ} iteratively. A solution is accepted if $\sum_{i,j}\left[ x_{\text{new}}(i,j)-x_{\text{old}}(i,j)\right]/L^2 \leqslant 10^{-10}$ in which $x(i,j)=n(i,j)/n_i$ and \textit{new} and \textit{old} refers to the updated and old solutions respectively. 
\subsection{Gaussian versus non-Gaussian random fields}
Let $h(\mathbf{r})$ be a single valued non-singular random field. A stochastic field is Gaussian if all its finite-dimensional probability distribution functions are Gaussian \cite{adler}. A necessary but not sufficient condition for Gaussian random field $h(\mathbf{r})$ is that its probability measure satisfies:
\begin{eqnarray}
\mathcal{P}\left\lbrace h \right\rbrace \equiv \frac{1}{\sigma\sqrt{2\pi}}e^{-\frac{h^2}{2\sigma^2}},
\end{eqnarray}
where $\sigma$ is the standard deviation. The local curvature at position $\mathbf{r}$ and at scale $b$
\begin{eqnarray}\label{local curvature}
C_b(\mathbf{r}) = \sum_{m=1}^M \left[ h(\mathbf{r}+ b\mathbf{e}_m) - h(\mathbf{r}) \right]
\end{eqnarray}
and the higher moments of $C_b$ are another measures to check the possible deviation of the random fluctuations from the Gaussian
distribution \cite{kondevpre}. In Eq. (\ref{local curvature}) the offset directions $\left\lbrace\mathbf{e}_1,\dots,\mathbf{e}_M\right\rbrace$ are a fixed set of vectors whose $\sum_{m=1}^M \mathbf{e}_m =0$. For a Gaussian stochastic field $h(\mathbf{r})$, the distribution of the local curvature $\mathcal{P}\left\lbrace C_b(\mathbf{r}) \right\rbrace$ is Gaussian and the first and all the other odd moments of $C_b$ are manifestly vanish since the random field has up/down symmetry $h(\mathbf{r})\longleftrightarrow -h(\mathbf{r})$. Obviously, for the Gaussian random fields the fourth moment satisfies:
\begin{eqnarray}\label{fourth moment}
\frac{\langle C_b^4 \rangle }{\langle C_b^2 \rangle ^2} = 3.
\end{eqnarray}
One should be careful about systematic deviations from the relation in Eq. (\ref{fourth moment})
that can occur for non-Gaussian random fluctuations. On the other hand, if a given random field contains hilltops and sharp valleys, $\langle C_b^3 \rangle \neq 0$ is a signature of skewness in the probability distribution functions \cite{kondevpre,hosseinabadi2014}. 
\begin{figure}[t]
\begin{center}
\includegraphics[width=8cm,clip]{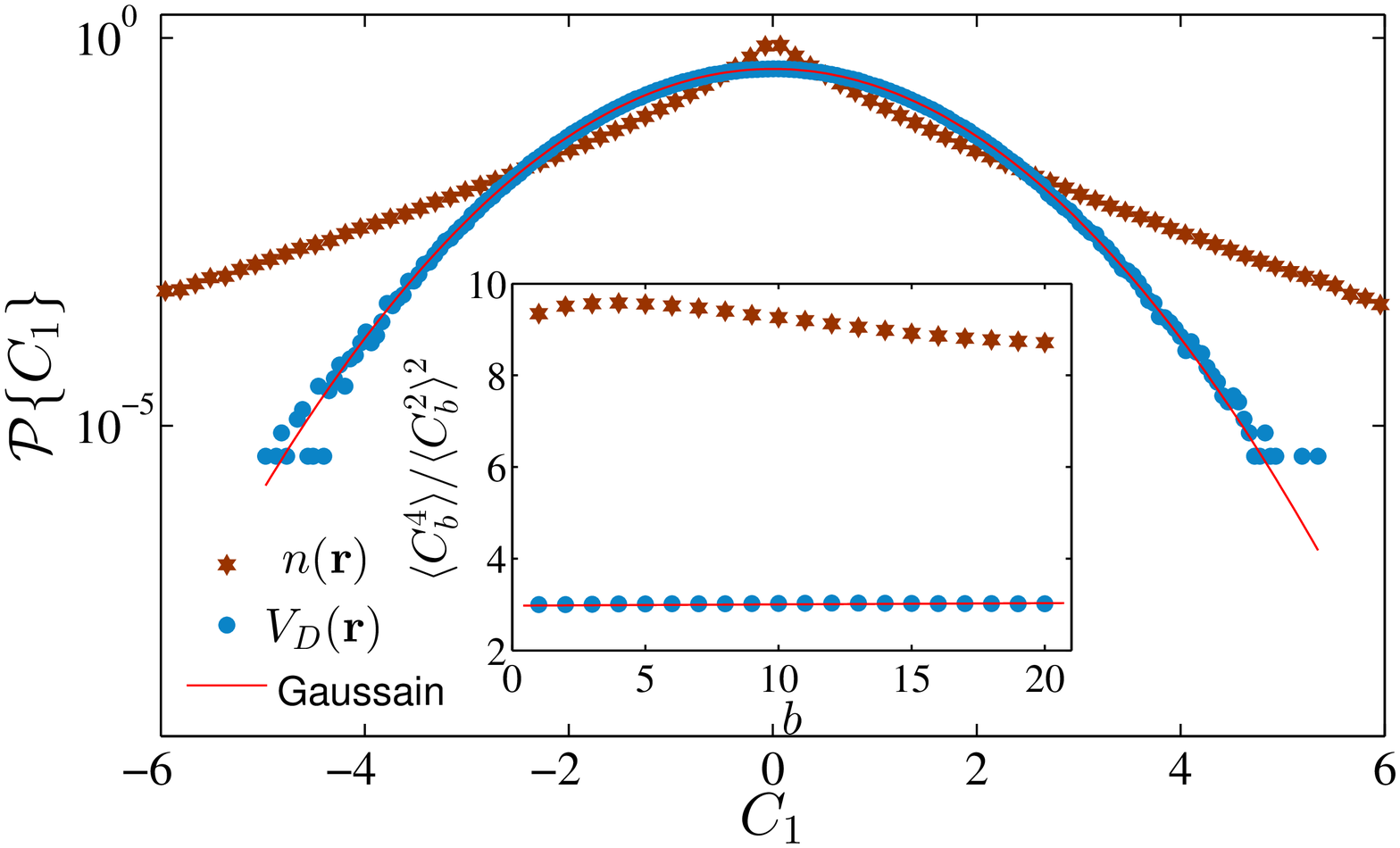}
\end{center}
\caption{(Color online) The probability distribution function $\mathcal{P}\lbrace C_b \rbrace$ ($b=1$) for the disorder potential $V_D(x,y)$ and the carrier density distribution $n(x,y)$ in semi-log scale. (Inset) the fourth moment $\frac{\langle C_b^4 \rangle }{\langle C_b^2 \rangle ^2}$ of the local curvature for $V_D$ and $n$. The prediction for a Gaussian surfaces is $3$.}
\label{figur3}
\end{figure}
As can be seen in Fig. (\ref{figur2}) the probability distribution function of the disorder potential $h=V_D$ is Gaussian and $\log \mathcal{P}\lbrace h \rbrace \propto -h^2$. In fact, this is one of the main consequences of the central limit theorem. The central limit theorem states that the probability distributions of sample sum of a sufficiently large number of independent random variables, with common probability density function with finite mean and variance, tend to be close to the normal distribution, regardless of the underlying distribution. As mentioned in Eq. \ref{VDis}, the disorder potential $V_D$ is the two dimensional Coulomb potential in the
graphene plane generated by an effective two dimensional uncorrelated random distribution,
$C(\mathbf{r})$. Therefore, we expect $V_D$ to be Gaussian because the integration Eq. \ref{VDis} is a linear functional of the Gaussian noise $C(\mathbf{r})$ and it is clear that adding independent mean zero Gaussian random variables with $\langle C(\mathbf{r}) \rangle = 0$ and $\langle C(\mathbf{r})C(\mathbf{r}^\prime) \rangle \propto \delta (\mathbf{r}^\prime - \mathbf{r})$ gives a Gaussian variable with $\langle V_D(\mathbf{r}) \rangle = 0$. We have also looked at the distribution function $\mathcal{P}\lbrace C_b\rbrace$ and the fourth moment (Eq. (\ref{fourth moment})) of the local curvature in disorder potential $V_D$. Figure (\ref{figur3}) shows that the distribution function for the local curvature for $V_D$ is Gaussian and the fourth moment of the local curvature for $V_D$ obeys the prediction which should be $3$ for a Gaussian surfaces. 
The probability distribution function $\mathcal{P}\lbrace h \rbrace$ of the carrier density $n(x,y)$ and the local curvature $\mathcal{P}\left\lbrace C_b(\mathbf{r}) \right\rbrace$ as well as the the fourth moment of the local curvature for $V_D$, are depicted in Figs. (\ref{figur2}) and (\ref{figur3}), respectively which exhibits non-Gaussian behaviors. This readily shows that the random field $n(\textbf{r})$ is non-Gaussian.
\subsection{Local and global roughness exponents}
\begin{figure}[t]
\begin{center}
\includegraphics[width=8cm,clip]{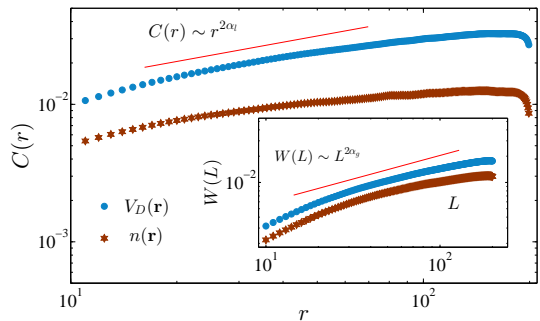}
\end{center}
\caption{(Color online) Log-Log plot of two point correlation function $C(r)$ as a function of $r$. The slope of this plot corresponds to local roughness exponent $\alpha_l$. (Inset) Log-Log plot of the profile width
$W(L)$ with respect to the window size $L$. The slope of this plot corresponds to global roughness exponent $\alpha_g$. 
}
\label{figur4}
\end{figure}

We should now calculate the exponents of the random field $n(\textbf{r})$ which has been sown to be non-Gaussian. We also calculate the exponents for $V_D$ for comparison. The scaling behavior of the two point correlation function of the random field $h$ was defined in Eq. (\ref{height-corr}). The measurement of local roughness exponent $\alpha_l$ can be obtained by a linear fit $C(r)$ with $r$ in $\log-\log$ scale. In Fig. (\ref{figur4}) we have plotted the scaling relation between $C(r)$ and $r$. In Table. I we report the scaling exponent $\alpha_l$ for disorder potential $V_D$ and carrier density distribution $n$. We have also computed the total variance $W(L)$, from which the exponent $\alpha_g$ is extracted using a scaling form Eq. (\ref{total variance}). The results are given in Fig. (\ref{figur4}) and our measurements of the scaling exponent $\alpha_g$ are reported in Table. I. From table I we see that the exponents $\alpha_l$ and $\alpha_g$ (see Table. I)) are the same within statistical errors as in the case of the mono-fractal random medium.\\
A question may arise here: how can the exponents be the same for $V_D(\textbf{r})$ and $n(\textbf{r})$, despite the fact that the former is Gaussian and the later is not? To answer this, let us consider Hohenberg-Kohn theorem according to which there is a one to one correspondence between the ground state charge density of a quantum system (here $n(\textbf{r})$) and the external potential (here $V_D$). This can be expressed by the relation $V_D=V_D[n]$ which may be a non-local function. Therefore the characteristic level lines of $V_D$ results in the same level lines for $n$ and the statitics are similar. Now consider the probability measure of them, i.e. $P(V_D)$ and $P(n)$. The mentioned relation implies the following equation: 
\begin{eqnarray}
P(V_D)\text{d}V_D=P(n)\text{d}n
\end{eqnarray}
according to which we have $P(n)=\left( \text{d}V_D/\text{d}n\right) P(V_D)$. Note that the necessary condition for this relation is that the conditional probability function $P(n|V_D)$ be a narrow function of both $V_D$ and $n$. This implies that given that $P(V_D)$ is Gaussian, the function $P(n)$ may not, depending on the quantity $\text{d}V_D/\text{d}n$.
\begin{table}[h]
\begin{tabular}{ l | c | c }
\hline 
& $\alpha_l$ & $\alpha_g$ \\
\hline
$V_D(x,y)$ & $0.47\pm 0.03$ & $0.45\pm 0.02$ \\
$n(x,y)$ & $0.35\pm 0.03$ & $0.38\pm 0.02$ \\
\hline 
\end{tabular}
\caption{The best fit values of the scaling exponents $\alpha_l$ and $\alpha_g$ extracted
from the scaling laws of two point correlation function $C(r)$ and the total variance $W(L)$ for disorder potential $V_D$ and carrier density distribution $n$.}
\label{tab1}
\end{table}
\subsection{Loop correlation exponent}
\begin{figure}[t]
\begin{center}
\includegraphics[width=8cm,clip]{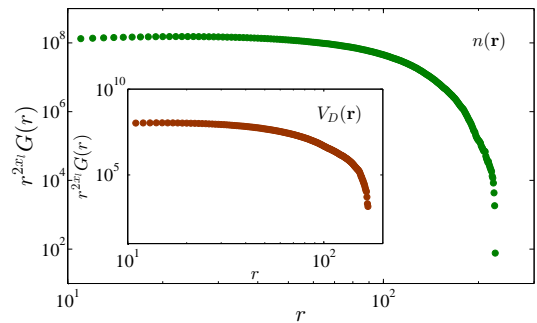}
\end{center}
\caption{(Color online) Log-Log plot of loop correlation function $r^{2x_l}G_l(r)$ as a function of $r$. 
}
\label{figur5}
\end{figure}

We now consider the loop correlation function $G_l(r)$, which is expected to behave like Eq. (\ref{loop correlation function}) for scale-invariant surfaces. To measure $x_l$, the most fundamental exponents of a given contour loop ensemble of the random profile $h$, we followed the numerical algorithm described in Ref. \cite{kondevpre}. For loops corresponding to mono-fractal rough interfaces with $\alpha =0$, based on exact results, $x_l = \frac{1}{2}$ \cite{kondevprl}. It has been checked in numerical simulations for the large classes of two dimensional Gaussian and non-Gaussian random fields that the relation $x_l = \frac{1}{2}$ is superuniversal which means that it is independent of the roughness exponent $\alpha$ \cite{kondevother,haghighi2011,hosseinabadi2012,hosseinabadi2014}. We measured the exponent $x_l$ from the power law dependence $G_l(r)$ with respect to $r$. In Fig. (\ref{figur5}) the $\log-\log$ plot of $r^{2x_l}G_l(r)$ as a function of $r$ has been indicated for the disorder potential $V_D$ and electron-hole distribution $n$. Our numerical test shows that $x_l = 0.5 \pm 0.1$ for the random fields $V_D$ and $n$. It is seen that it is the same as the reported value for the mono-fractal rough interfaces \cite{kondevprl}. The relatively large error bar comes from finite size effects. It is worth mentioning that the relation $x_l =\frac{1}{2}$ is valid for a non-Gaussian interface, i.e. $n$, as well as the Gaussian profile $V_D$ (see also Refs. \cite{hosseinabadi2014}). 
\subsection{Fractal dimensions}
We now present the detailed analysis of fractal properties of the mean level contour lines of the disorder potential $V_D$ and electron-hole distribution $n$. We used the self-similar properties of contour lines to measure fractal dimension of loops $D_f$ and the fractal dimension of all the contours $d$. 
\begin{table}[h]
\begin{tabular}{ l | c | c }
\hline 
& $V_D(x,y)$ & $n(x,y)$ \\
\hline
$D_f$ & $1.38\pm 0.02$ & $1.39\pm 0.01$ \\
$d$ & $1.80\pm 0.03$ & $1.80\pm 0.03$ \\
$\zeta/2$ & $0.91\pm 0.03$ & $0.90\pm 0.02$ \\
$\tau$ & $2.30\pm 0.02$ & $2.30\pm 0.01$ \\
\hline
\hline
$D_f(\tau-1)/\zeta$ & $0.99\pm 0.04$ & $1.00\pm 0.03$ \\
\hline
$D_f(\tau-3)/(2x_l-2)$ & $0.97\pm 0.21$ & $0.97\pm 0.20$ \\
\hline 
\end{tabular}
\caption{The best fit values of the scaling exponents $D_f$, $d$, $\zeta$ and $\tau$ for disorder potential $V_D$ and carrier density distribution $n$.}
\label{tab2}
\end{table} 
\subsubsection*{Length-radius scaling relation}
\begin{figure}[t]
\begin{center}
\includegraphics[width=8cm,clip]{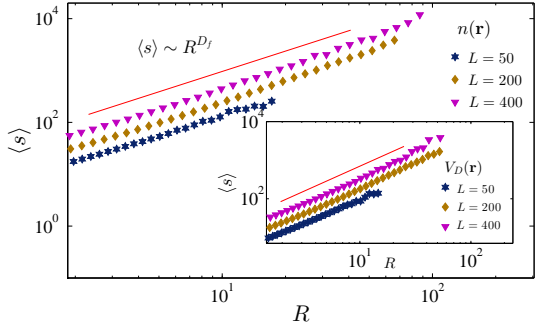}
\end{center}
\caption{(Color online) The power law scaling relation between the mean value of the loop length $\langle s \rangle$ and the gyration radius $R$ for
contour lines of the electron-hole distribution $n$ and the disorder potential $V_D$ (inset). 
}
\label{figur6}
\end{figure}

In order to evaluate the fractal dimension of the contour loops $D_f$ we used the scaling law between the mean value of the loop length $\langle s\rangle$ and its radius of gyration $R$ according to Eq. \ref{s-R}. For a given loop with $N$ discrete points $\lbrace \mathbf{r}_1,\dots,\mathbf{r}_N \rbrace$, the radius of gyration is defined by $R^2 = \frac{1}{N}\sum_{i=1}^N |\mathbf{r}_i - \mathbf{r}_c|^2$ where $\mathbf{r}_c=\frac{1}{N}\sum_{i=1}^N \mathbf{r}_i$ is the center of mass of the contour line. A plot picturing the scaling of the mean loop length as a function of the radius for the contour loop ensemble of the disorder potential $V_D$ and electron-hole distribution $n$ with different system size $L$, is shown in Fig. (\ref{figur6}). 
The straight line in the figure shows power law scaling with the fractal dimension exponent $D_f$. The scaling exponent $D_f$ is measured using the linear fit and chi-square test in the scaling regime ($5 < R < 100$). In Table II we report the fractal dimension of contours $D_f$ for the random profiles $V_D$ and $n$. It is not difficult to see that for a random field which is self affine with universal value of the loop correlation exponent $x_l=\frac{1}{2}$, the formula for the fractal dimension $D_f$ follows from $D_f = \frac{3-\alpha}{2}$ \cite{kondevprl}. In the case of the contour lines of the disorder potential $V_D$ and electron-hole distribution $n$, the fractal dimension of a contour line follows the formula of a mono fractal interfaces with roughness exponent $\alpha$, even in the case of random fields with non-Gaussian probability distributions, i.e. the electron-hole distribution $n$. 
\subsubsection*{Fractal dimension of all contours}
\begin{figure}[t]
\begin{center}
\includegraphics[width=8cm,clip]{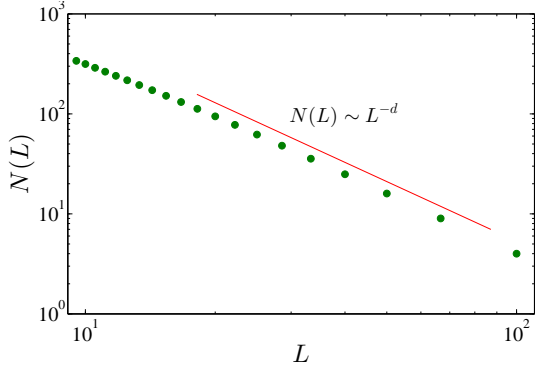}
\end{center}
\caption{(Color online) The fractal dimension of all contours: the log-log plot of the number of boxes $N(l)$ in terms of the box size $l$.
}
\label{figur6.5}
\end{figure}
Another interesting scaling exponent which describes the scaling properties of random profile $h$, is the fractal dimension of all contours. It is numerically showed that all disconnected loops in the contour loop ensemble of constant height on self-affine random profile $h(x,y)$, i.e. Fig. (\ref{figur1}), is also a self-similar fractal with fractal dimension $d=2-\alpha$ \cite{mandelbrot}. The fractal dimension of all contours in the mean level set can be found by box-counting method \cite{william}. The basic procedure is to cover the level set with the set of $l$-sized boxes, and then count the number of boxes $N(l)$ which are covering the level set. Then we do the same thing but using a smaller boxes. For a mono-fractal object the scaling law between the number of boxes $N(l)$ and the box size $l$ is
\begin{eqnarray}
N(l)\sim l^{-d},
\end{eqnarray}
where $d$ is the fractal dimension. The results of the fractal dimension analysis of all contours are given in Fig. (\ref{figur6.5}). In Table II, we report the best linear data fit to $\log N(l)$ with respect to $\log l$, yielding the fractal dimension of all contours $d$ for Gaussian profile $V_d$ and non-Gaussian random distribution $n$. Our numerical tests confirm the validity of the relation $d = 2-\alpha$ within our statistical errors. 
\subsubsection*{Cumulative distribution of areas}
\begin{figure}[t]
\begin{center}
\includegraphics[width=8cm,clip]{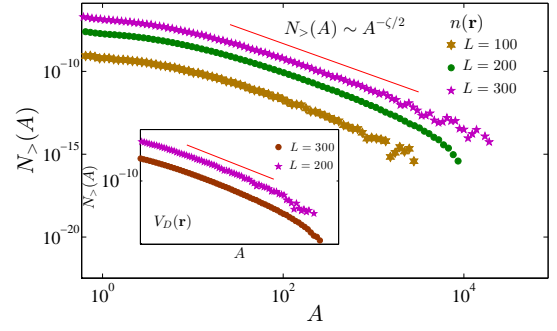}
\end{center}
\caption{(Color online) The scaling behaviour of the cumulative number of loops whose
area is greater than $A$ as a function of area for the electron-hole distribution $n$ and the disorder potential $V_D$ (inset).
}
\label{figur7}
\end{figure}
We will now consider in detail the procedure for extracting the scaling behavior of the cumulative distribution of the loop area $N_>(A)$. For the case of self-affine random field the number of contours with area greater than $A$ has the asymptotic scaling behavior of Eq. (\ref{scaling of cumulative dist}) with the scaling exponent $\zeta$. This can be seen in Fig. (\ref{figur7}) which illustrates the log-log plot of $N_>(A)$ versus $A$ for the disorder potential $V_D$ and electron-hole distribution $n$. The slope of such a plot determines the exponent $\zeta$. The measured value of $\zeta$ has been reported in Table II. Note that following Ref. \cite{kondevpre}, the exponent $\zeta$ is related to the fractal dimension of all contours by $\zeta = d/2$. Our numerical results are consistent with $\zeta = 1-\alpha/2$. 
\begin{figure}[t]
\begin{center}
\includegraphics[width=8cm,clip]{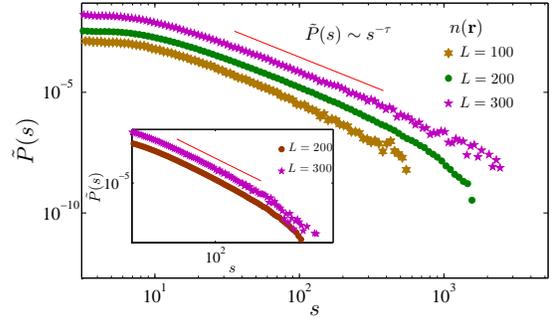}
\end{center}
\caption{(Color online) The scaling behaviour of the loop-size distribution $\tilde{P}(s)$ for the electron-hole distribution $n$ and the disorder potential $V_D$ (inset).
}
\label{figur8}
\end{figure}
\subsection{Length distribution exponent}
Let us now focus on the probability distribution function of the loop length $\tilde{P}(s)$ which follows the scaling law Eq. (\ref{pdf of contour size}) with the loop exponent $\tau$. As shown in Fig. (\ref{figur8}), we presented the log-log plot of $\tilde{P}(s)$ versus $s$. The length distribution exponent $\tau$ can
be measured numerically from the power-law scaling regime which is evident over two
decades in loop length ($10 < s < 1000$). The numerical results of the loop distribution exponent $\tau$ for the disorder potential $V_D$ and electron-hole distribution $n$ are
reported in Table. II. We also checked numerically the consistency of the results for different finite system size. It is also worth noting that the exponent $\tau$ even for a non-Gaussian random profile ($n$) satisfies 
$\tau = 1+(4-2\alpha)/(3-\alpha)$. 
We emphasize that, for the given numerical values of the scaling exponents $x_l$, $D_f$, $d$, $\zeta$ and $\tau$, which are summarized in Table II, it is straightforward to check that
according to Eqs. (\ref{hyper1}) and (\ref{hyper2}), the hyper scaling relations, are valid for both Gaussian random profile $V_D$ and non-Gaussian random distribution $n$. From Eqs. (\ref{hyper1}) and (\ref{hyper2}) it follows that $D_f(\tau-1)/\zeta$ and $D_f(\tau-3)/(2x_l-2)$ are equal to one. It is seen that our results are in a good agreement with the
theoretical prediction. 
\section{Conclusion}
In this paper we considered the zero-temperature Thomas-Fermi-Dirac (TFD) theory for graphene at the Dirac point. Based on some stochastic analysis we obtained the probability measure of the ground state career density for small interactions and small impurity concentrations. We argued that in vicinity of the Dirac point the density fluctuations increase unboundedly, leading to a new phase at which large charge inhomogeneities arise, i.e. EHPs. Since the mentioned calculations are not valid for all range of interactions and impurity concentrations, we solved the TFD equation numerically and over $6\times 10^3$ samples of various sizes were generated. As argued analytically, we observed power-law behaviors for the ground state charge density $n$ which is expected from the scale invariance of the equation governing $n$. When viewed as random field surface, the impurity potential field $V_D$ was found to be Gaussian as expected, whereas the ground state charge density was not. The evidence for this is the probability distribution of them which is Gaussian for the former and non-Gaussian for the later. We precisely analyzed the various exponents of the system. Local and global roughness exponents are found to be the same for both $n$ and $V_D$ which is the signature of mono-fractal behavior of the surface. Loop correlation exponent is also found to be equal to the super-universal value $1/2$ for both. Various fractal dimensions and length distribution exponent are also reported and found to be the same for $n$ and $V_D$. Although not a Gaussian random field, the charge density is found interestingly to satisfy the Kondev scaling relations. 

\end{document}